\documentclass[a4paper,11pt]{article}
\usepackage{pos}
\usepackage{bm}
\usepackage{physics}
\usepackage{slashed}
\usepackage{mathtools}
\usepackage{tikz-cd}
\usepackage{feynmf} 
\title{Dilaton Sum Rules of Gravitational Form Factors in QCD at Order $\alpha_s$}
\ShortTitle{Dilaton Sum rules in QCD and anomaly form factors}

\author*[a,b,c]{Claudio Corian\`o}
\author[a]{Stefano Lionetti}
\author[a,b]{Dario Melle}
\author[a]{Leonardo Torcellini}

\affiliation[a]{Dipartimento di Matematica e Fisica ``Ennio De Giorgi'', Universit\`a del Salento and INFN Sezione di Lecce,\\
Via per Arnesano, 73100 Lecce, Italy}
\affiliation[b]{National Center for HPC, Big Data and Quantum Computing, Italy}
\affiliation[c]{CNR Nanotec, Lecce, Italy}

\emailAdd{claudio.coriano@unisalento.it}

\FullConference{Corfu Summer Institute 2025\\
Corfu, Greece}

\abstract{%
We formulate a partonic description of hadronic gravitational form factors within QCD, focusing on the three-point function of the energy-momentum tensor and two gluon currents. Despite the lack of exact conformal symmetry in QCD, the correlator may be organized around the conformal limit through momentum-space CFT methods, suitably adjusted for gauge-fixing effects. This yields a tensor decomposition into spin-2, spin-1, and spin-0 sectors, with the spin-0 contribution governed by the conformal anomaly. The corresponding anomaly form factor satisfies a mass-independent dispersive sum rule and allows a dilaton-like interpretation. In the light-cone limit, this term and an additional traceless structure become dominant, indicating an effective anomaly-mediated description relevant to hadronic gravitational form factors.}

\begin{document}
\maketitle

\section{Introduction}

Gravitational form factors (GFFs) provide a compact description of hadron structure in terms of matrix elements of the QCD energy-momentum tensor. They encode momentum fractions, total and orbital angular momentum, and the mechanical properties of hadrons such as pressure and shear forces. Experimentally they are not accessed through dynamical gravity, but through hard exclusive processes, most notably deeply virtual Compton scattering (DVCS), where generalized parton distributions (GPDs) enter the factorized amplitude and connect measurable observables to moments of the stress energy tensor \cite{Ji:1998pc,Radyushkin:1997ki,Ji:1996ek,Bhattacharya:2022xxw,Bhattacharya:2023wvy}.\\
The relevant hard subprocesses can be analyzed through correlators involving insertions of this tensor. The basic subleading interaction that we are going to consider is the non-Abelian three-point function $TJJ$, with one stress energy tensor and two gauge currents, shown as an insertion in Fig.1 (right). This correlator is the natural arena in which the conformal anomaly, its dispersive representation and its interpolation by effective scalar degrees of freedom can be studied perturbatively \cite{Armillis:2009pq,Armillis:2010qk,Coriano:2018bbe,Coriano:2024qbr}. The purpose of the present contribution is to summarize this picture and to make explicit the equations underlying the discussion.\\
 In this contribution, briefly, we first recall the relation between DVCS, generalized parton distributions (GPDs) and GFFs. We then discuss the curved-space formulation of QCD and the $TJJ$ correlator, highlighting how momentum-space CFT methods  \cite{Bzowski:2013sza,Bzowski:2018fql} \cite{Coriano:2020ees} can still be exploited despite the explicit breaking of conformal symmetry by the gauge-fixing sector. Finally, we describe the emergence of a conformal-anomaly form factor from the general structure of the perturbative correlator, its mass-independent sum rule, and the light-cone dominance of the anomaly and of an additional traceless structure (also pole-like) that survives in the same limit. The traceless structure was first identified in \cite{Giannotti:2008cv} as a plasmon mode, which is also present in the non-Abelian analysis. Details of this work can be found in 
\cite{GRSUMRULES-FINALE3MOD} and in \cite{Coriano:2024qbr}. Similar analysis, in the context of chiral corrrelators and the gravitational chiral anomaly can be found in \cite{Coriano:2025ceu}. 

\section{DVCS, generalized parton distributions and gravitational form factors}

The standard phenomenological entry point is DVCS, where one probes non-forward matrix elements of quark and gluon operators on the light cone. In the generalized Bjorken limit the amplitude factorizes into perturbative coefficient functions and nonperturbative GPDs \cite{Ji:1998pc,Radyushkin:1997ki}. \\
The Mellin moments of the non-forward parton distributions reconstruct matrix elements of the QCD stress energy tensor and therefore the GFFs of the hadron. In particular, Ji's sum rule provides the best-known example of this connection, relating the total partonic angular momentum to the forward limits of suitable form factors \cite{Ji:1996ek},
\begin{equation}
J_q+J_g=\frac12 \left[A_q(0)+B_q(0)+A_g(0)+B_g(0)\right]=\frac12 .
\label{eq:Ji}
\end{equation}

At the hadronic level one considers matrix elements of the form
\begin{equation}
\mel{p',s'}{T^{\mu\nu}(0)}{p,s},
\label{eq:emtme}
\end{equation}
whose Lorentz decomposition defines the quark and gluon GFFs. For a spin-$1/2$ target one may write
\begin{align}
\mel{p',s'}{T^{\mu\nu}_{q,g}(0)}{p,s} = \bar u(p',s') \Bigg[& A_{q,g}(t)\, \gamma^{(\mu} P^{\nu)}
+ B_{q,g}(t)\, \frac{P^{(\mu} i\sigma^{\nu)\rho}\Delta_\rho}{2M} \notag\\
&+ D_{q,g}(t)\,\frac{\Delta^\mu\Delta^\nu-g^{\mu\nu}\Delta^2}{4M}
+ \bar C_{q,g}(t)\,M g^{\mu\nu}\Bigg] u(p,s),
\label{eq:gffdecomp}
\end{align}
with $P=(p+p')/2$, $\Delta=p'-p$ and $t=\Delta^2$. The form factors $A$, $B$ and $D$ encode, respectively, momentum flow, angular momentum information and mechanical properties. The experimental access to these form factor is indirect, and takes place via specific nonlocal correlators on the light cone. For example, in the quark case, these take the form 
\begin{equation}
F^q(x,\xi,t)=\frac12\int \frac{dz^-}{2\pi}
\,e^{ixP^+z^-}
\mel{p'}{\bar\psi\!\left(-\frac{z}{2}\right)\gamma^+\mathcal{W}\psi\!\left(\frac{z}{2}\right)}{p}\Bigg|_{z^+=0,\,\bm z_\perp=0},
\label{eq:gpddef}
\end{equation}
parametrized in terms of the quark GPDs $H^q$ and $E^q$, where $\mathcal{W}$ dentes the gauge link. A similar expression holds for the gluon sector. The first Mellin moments of the GPDs yield
\begin{align}
\int_{-1}^{1} dx\, x\, H^q(x,\xi,t) &= A_q(t)+\xi^2 D_q(t),
\label{eq:mellin1}\\
\int_{-1}^{1} dx\, x\, E^q(x,\xi,t) &= B_q(t)-\xi^2 D_q(t),
\label{eq:mellin2}
\end{align}
and similarly for the gluon sector. Therefore the hard exclusive DVCS amplitude provides indirect access to matrix elements of the QCD stress tensor, after integration over one of its scaling variables $(x,\xi)$. The factorized structure of the non-forward amplitude can be represented symbolically as
\begin{equation}
\mathcal{A}_{\rm DVCS} \sim \int dx\, T_H(x,\xi,Q^2,\mu^2)\, F(x,\xi,t;\mu^2),
\label{eq:dvcsfact}
\end{equation}
where $T_H$ is the perturbative kernel. Differently from inclusive processes, in DVCS the factorization accurs at amplitude level. \\
The insertion of the energy--momentum tensor into the hard subprocess allows the partonic amplitude to be interpreted as the short-distance kernel underlying the hadronic gravitational form factors (GFFs). It is in this framework that the $TJJ$ correlator becomes relevant. In Fig.~1 we show representative leading-order and next-to-leading-order (NLO) contributions to the pion GFF, in Fig.~2 typical partonic contributions, while the structure of the corresponding perturbative corrections in the proton case is displayed in Fig.~3. An analysis of these contributions has been presented in~\cite{Tong:2022zax}. In the following, we focus on a specific class of NLO corrections in which the $TJJ$ vertex, and with it the conformal anomaly, enters explicitly in the description.

\begin{figure}[t]
\begin{center}
\includegraphics[scale=0.4]{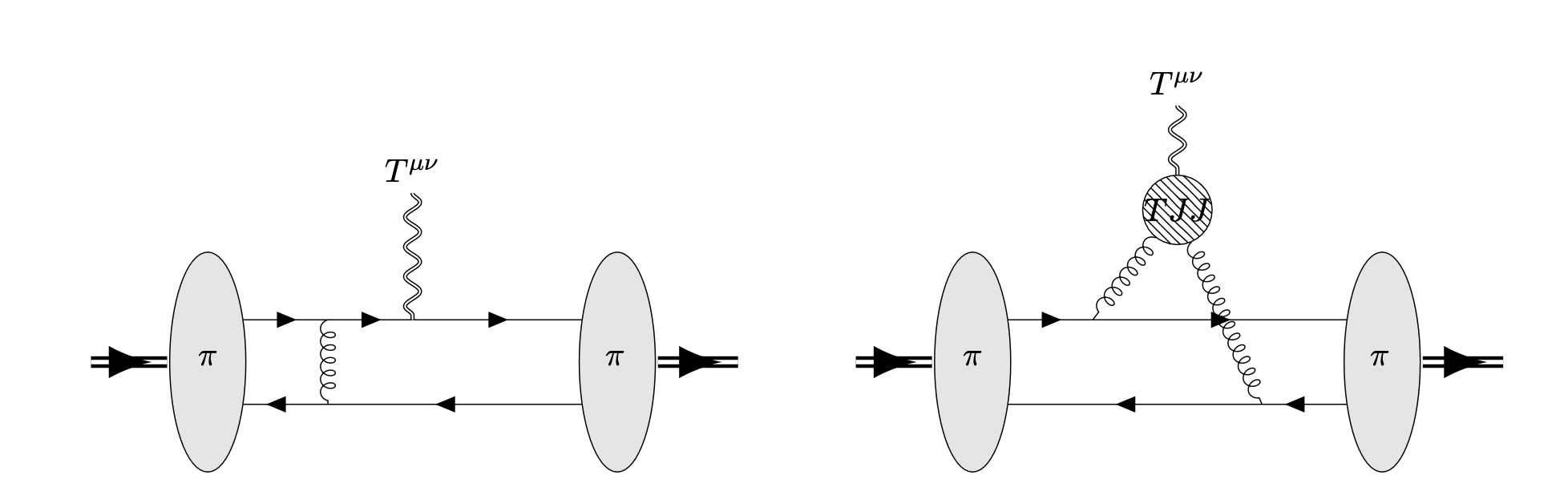}
 \caption{Leading (left) and NLO contributions to the GFF of the pion.}
\label{expansion}
\end{center}
\end{figure}

\section{QCD in curved spacetime and the trace anomaly}
To investigate the anomaly sector, one couples QCD to an external metric and treats the latter as a classical source for the full QCD energy--momentum tensor, as shown in \cite{Coriano:2024qbr}. Embedding the QCD action in a curved background is a natural step in the study of the conformal constraints satisfied by the theory. The analysis presented in 
our works follows closely earlier perturbative investigations of correlators in free-field theories \cite{Coriano:2018bbe,Armillis:2009pq}\cite{Giannotti:2008cv}  and provides a bridge between the general conformal field theory (CFT) methods to determine the correlation functions of 3-point functions and their explicit realization in a Lagrangian framework. While many of the steps introduced needed in order to solve the conformal constraints are identical to the Abelian case \cite{Bzowski:2013sza,Bzowski:2018fql}, in the case of a non-abelian theory and in the factorization picture, there are are several changes \cite{GRSUMRULES-FINALE3MOD}. These are due to the presence of a gauge fixing sector and of BRST constraints, which render the approach more involved. Indeed, in the perturbative non-Abelian case, however,  the presence of the gauge-fixing and ghost sectors, require a refinement of the standard momentum-space CFT analysis  \cite{Coriano:2024qbr}.\\
We start from the generating functional of QCD, obtained integrating over the quark, the gauge and
 the ghost and antighost fields $(c,\bar c)$.
\begin{equation}
Z[g,A]=\int \mathcal{D}A\mathcal{D}\psi\, \mathcal{D}\bar\psi\,\mathcal{D}c\,\mathcal{D}\bar c\,
\exp\Bigl(iS_{\rm QCD}[g,A,\psi,\bar\psi,c,\bar c]\Bigr),
\label{eq:partition}
\end{equation}
from which one derives the expectation value of the stress energy tensor
\begin{equation}
\expval{T^{\mu\nu}(x)}=\frac{2}{\sqrt{g(x)}}\frac{\delta \ln Z[g,A]}{\delta g_{\mu\nu}(x)}.
\label{eq:tdef}
\end{equation}
Functional differentiation of this relation with respect to the gauge source then generates correlators with one or more insertions of gauge currents. The lowest nontrivial one is the three-point function
\begin{equation}
\Gamma^{\mu\nu\alpha\beta ab}(z,x,y)=\left.\frac{\delta^2}{\delta A^a_\alpha(x)\delta A^b_\beta(y)}\expval{T^{\mu\nu}(z)}\right|_{A=0},
\label{eq:tjjdef}
\end{equation}
which is precisely the $TJJ$ vertex relevant for the present discussion \cite{Armillis:2009pq,Armillis:2010qk}. The correlator plays a key role in the coupling of Standard Model matter to gravity. Analysis of this 
interaction have been presented in the past in the electroweak theory \cite{Coriano:2011zk}.

\subsection{The reconstruction of the quark and gluon sectors in QCD}

The perturbative analysis requires the QCD stress energy tensor. In the gauge-fixed theory it is obtained by combining the field-strength, fermionic, gauge-fixing and ghost contributions,
\begin{equation}
T_{\mu\nu}=T^{\rm f.s.}_{\mu\nu}+T^{\rm ferm.}_{\mu\nu}+T^{\rm g.f.}_{\mu\nu}+T^{\rm gh}_{\mu\nu}.
\label{eq:emtsplit}
\end{equation}
The field-strength contribution is
\begin{equation}
T^{\rm f.s.}_{\mu\nu}=-\eta_{\mu\nu}\,\frac14 F^a_{\rho\sigma}F^{a\rho\sigma}+F^a_{\mu\rho}{F^a_{\nu}}^{\rho},
\label{eq:emtfs}
\end{equation}
while the gauge-fixing and ghost sectors take the form
\begin{align}
T^{\rm g.f.}_{\mu\nu}={}&-\frac{1}{\xi}\left[A^a_{\nu}\,\partial_{\mu}(\partial\!\cdot\!A^a)+A^a_{\mu}\,\partial_{\nu}(\partial\!\cdot\!A^a)\right]
+\frac{1}{\xi}g_{\mu\nu}\left[-\frac12(\partial\!\cdot\!A^a)^2+\partial^{\rho}\bigl(A^a_{\rho}\,\partial\!\cdot\!A^a\bigr)\right],
\label{eq:emtgf}\\[2mm]
T^{\rm gh}_{\mu\nu}={}&-\partial_{\mu}\bar c^a D^{ab}_{\nu}c^b-\partial_{\nu}\bar c^a D^{ab}_{\mu}c^b+g_{\mu\nu}\,\partial^{\rho}\bar c^a D^{ab}_{\rho}c^b.
\label{eq:emtgh}
\end{align}
After the inclusion of the fermion sector, the full expression of the QCD energy-momentum tensor becomes
\begin{align}
T_{\mu\nu}={}&-g_{\mu\nu}\,{\cal L}_{\rm QCD}-F^a_{\mu\rho}{F^a_{\nu}}^{\rho}-\frac{1}{\xi}g_{\mu\nu}\,\partial^{\rho}\bigl(A^a_{\rho}\partial^{\sigma}A^a_{\sigma}\bigr)
+\frac{1}{\xi}\Bigl(A^a_{\nu}\partial_{\mu}(\partial^{\sigma}A^a_{\sigma})+A^a_{\mu}\partial_{\nu}(\partial^{\sigma}A^a_{\sigma})\Bigr)
\nonumber\\
&+\frac{i}{4}\Bigl[\bar\psi\gamma_{\mu}(\overrightarrow\partial_{\nu}-igT^aA^a_{\nu})\psi-\bar\psi(\overleftarrow\partial_{\nu}+igT^aA^a_{\nu})\gamma_{\mu}\psi
+\bar\psi\gamma_{\nu}(\overrightarrow\partial_{\mu}-igT^aA^a_{\mu})\psi
\nonumber\\
&\hspace{1.7cm}-\bar\psi(\overleftarrow\partial_{\mu}+igT^aA^a_{\mu})\gamma_{\nu}\psi\Bigr]
+\partial_{\mu}\bar c^a\bigl(\partial_{\nu}c^a-gf^{abc}A^c_{\nu}c^b\bigr)
+\partial_{\nu}\bar c^a\bigl(\partial_{\mu}c^a-gf^{abc}A^c_{\mu}c^b\bigr).
\label{eq:EMTfull}
\end{align}
Here $\xi$ is the gauge-fixing parameter, $A_\mu^a$ are the gauge fields, $F$ is their field strength. The normalization of the non-Abelian stress tensor is chosen in agreement with \cite{Armillis:2010qk}. Its derivation follows from the embedding of the gauge-fixed QCD action in a curved background, with definition
\begin{equation}
T^{\mu\nu}_{\rm QCD}\equiv \frac{2}{\sqrt{-g}}\frac{\delta S_{\rm QCD}}{\delta g_{\mu\nu}}.
\label{eq:emtcurved}
\end{equation}
The decomposition in \eqref{eq:EMTfull} is important for two reasons. First, it makes explicit that the gluon, ghost and gauge-fixing sectors all contribute to the three-point function, and therefore to the reconstruction of the complete QCD anomaly vertex. Second, it clarifies why the quark and gluon sectors behave differently from the viewpoint of conformal symmetry: the fermion sector is closer to the CFT solution, whereas the gauge-fixed bosonic sector carries the explicit breaking associated with BRST quantization. \\
The anomaly is generated by the renormalization procedure. It is contained in the operator relation \cite{Collins:1976yq,Adler:1976zt,Chanowitz:1972vd}
\begin{equation}
\eta_{\mu\nu}\expval{T^{\mu\nu}}=-\frac{\beta(g_s)}{2g_s}\,F^a_{\rho\sigma}F^{a\rho\sigma},
\label{eq:traceanom}
\end{equation}
valid in the massless theory, with one-loop QCD beta function
\begin{equation}
\beta(g_s)=\frac{g_s^3}{16\pi^2}\,\frac{-11C_A+2n_f}{3}.
\label{eq:betafun}
\end{equation}
The same relation shows why QCD is only classically conformal before gauge fixing and in the massless limit.  From the functional point of view one may in principle analyze a whole hierarchy of correlators $TJ^n$, with $n=2,3,4,\ldots$, obtained by differentiating Eq.~\eqref{eq:tdef} repeatedly with respect to the gauge source. In the non-Abelian case this hierarchy is relevant because the operator $F^2$ contains cubic and quartic gluonic terms. However, for the determination of the anomaly coefficient itself the $TJJ$ vertex is already sufficient: after tracing over the graviton indices, its coefficient of the distinguished two-gluon tensor structure reproduces the full one-loop beta function.
If fermion masses are included, extra explicit breaking terms appear and one must disentangle them from the genuine anomaly contribution. At the level of the three-point function this is encoded in the Ward identity
\begin{equation}
\eta_{\mu\nu}\,\Gamma^{\mu\nu\alpha\beta ab}=\mathcal{A}^{\alpha\beta ab}+\expval{{T^\mu}_\mu J^{\alpha a}J^{\beta b}},
\label{eq:traceWI}
\end{equation}
where $\mathcal{A}^{\alpha\beta ab}$ denotes the contribution generated by differentiating the anomaly functional with respect to the gauge fields. The first term is fixed by the anomaly functional and is mass independent, whereas the second contains the explicit insertion of the trace of the stress energy tensor and therefore carries the dependence on quark masses and on the details of the broken phase. This distinction is central when identifying the genuine anomaly form factor away from the strict conformal point, and needs to be taken into account in the analysis of the sum rule. It is conceptually the same separation encountered in the electroweak theory when one isolates the genuine anomalous dilaton--gauge--gauge coupling from ordinary loop-induced scale breaking proportional to the Higgs vacuum expectation value or to explicit fermion masses \cite{Coriano:2011zk}.

At leading order the anomaly contribution differentiated twice with respect to the gluon background takes the form
\begin{equation}
\mathcal{A}^{\alpha\beta ab}(p_1,p_2)=\frac{g_s^2}{48\pi^2}(11C_A-2n_f)\,\delta^{ab}
\,u^{\alpha\beta}(p_1,p_2),
\label{eq:anomAA}
\end{equation}
with
\begin{equation}
u^{\alpha\beta}(p_1,p_2)=(p_1\!\cdot\! p_2)g^{\alpha\beta}-p_2^\alpha p_1^\beta,
\label{eq:utensor}
\end{equation}
a tensor proportional to the Fourier transform of $F^2$ and transverse with respect to both external gluon momenta in the on-shell limit
\begin{eqnarray}
&&u^{\alpha\beta a_1 a_2}(p_1,p_2, a_1, a_2) = -\frac{1}{4}\int\,d^4x\,\int\,d^4y\ e^{ip_1\cdot x + i p_2\cdot y}\ 
\frac{\delta^2 \{F^a_{\mu\nu}F^{a\mu\nu}(0)\}} {\delta A_{\alpha}^{a_1}(x) A_{\beta}^{a_2}(y)}\vline_{A^a=0} \,
\label{locvar}
\end{eqnarray}

 $a$,  $a_1,a_2$ are colour indices.

\section{Momentum-space CFT and the role of gauge fixing}

The $TJJ$ correlator can be organized using the techniques of CFT in momentum space \cite{Coriano:2013jba,Bzowski:2013sza,Bzowski:2018fql,Coriano:2018bbe,Coriano:2017mux,Coriano:2018bsy}. In a genuine conformal theory, three-point functions of currents and stress energy tensors are constrained by conformal Ward identities and can be decomposed into transverse-traceless, longitudinal and trace sectors. The elementary projectors are
\begin{equation}
\pi^{\mu}{}_{\alpha}(p)=\delta^{\mu}{}_{\alpha}-\frac{p^\mu p_\alpha}{p^2},
\label{eq:piprojector}
\end{equation}
and the transverse-traceless projector acting on a rank-two tensor is
\begin{equation}
\Pi^{\mu\nu}{}_{\alpha\beta}(p)=\pi^{(\mu}{}_{\alpha}(p)\pi^{\nu)}{}_{\beta}(p)-\frac{1}{d-1}\pi^{\mu\nu}(p)\pi_{\alpha\beta}(p),
\label{eq:TTprojector}
\end{equation}
with $d$ denoting the spacetime dimensions.
In the same language one can decompose a generic tensor according to
\begin{equation}
\delta_{\mu(\alpha}\delta_{\beta)\nu}=
\Pi_{\mu\nu\alpha\beta}(p)+\mathscr{T}_{\mu\nu(\alpha}(p)p_{\beta)}+\frac{1}{d-1}\pi_{\mu\nu}(p)g_{\alpha\beta},
\label{eq:splitprojector}
\end{equation}
where $\mathscr{T}_{\mu\nu\alpha}$ denotes a longitudinal projector with pivot the four momentum of the graviton. This decomposition is extremely convenient when organizing the partonic amplitude.\\
QCD is not conformal as a full interacting theory, but even at tree level  the gauge-fixed action contains a gauge-fixing term and the corresponding ghost sector breaking the conformal symmetry. The quark contribution to $TJJ$ is therefore closer to the standard conformal solution than the gluon sector, whose off-shell Ward identities are replaced by Slavnov-Taylor relations. We emphasize that, despite these complications, the momentum-space CFT framework still provides the appropriate language for identifying the physical tensor structures and, in particular, the scalar sector that carries the anomaly. 
This analysis, covering the extension of the reconstruction of \cite{Bzowski:2013sza,Bzowski:2018fql} to gauge fixed non-Abelian theories, is contained in \cite{GRSUMRULES-FINALE3MOD} with open indices. 
Radiative corrections of the trace anomaly in QCD have been discussed in \cite{Hatta:2018sqd}.

\begin{figure}[t]
\begin{center}
\includegraphics[scale=0.5]{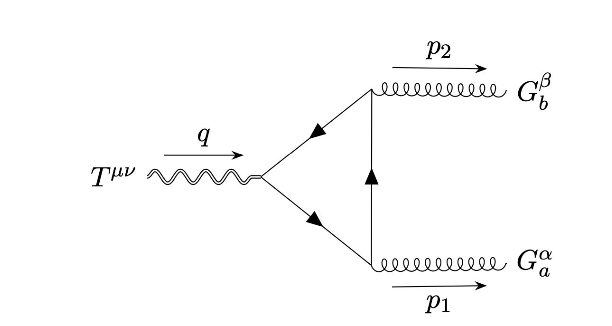}
\includegraphics[scale=0.6]{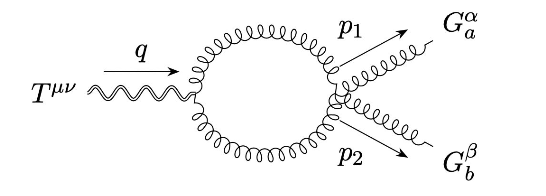}
\includegraphics[scale=0.14]{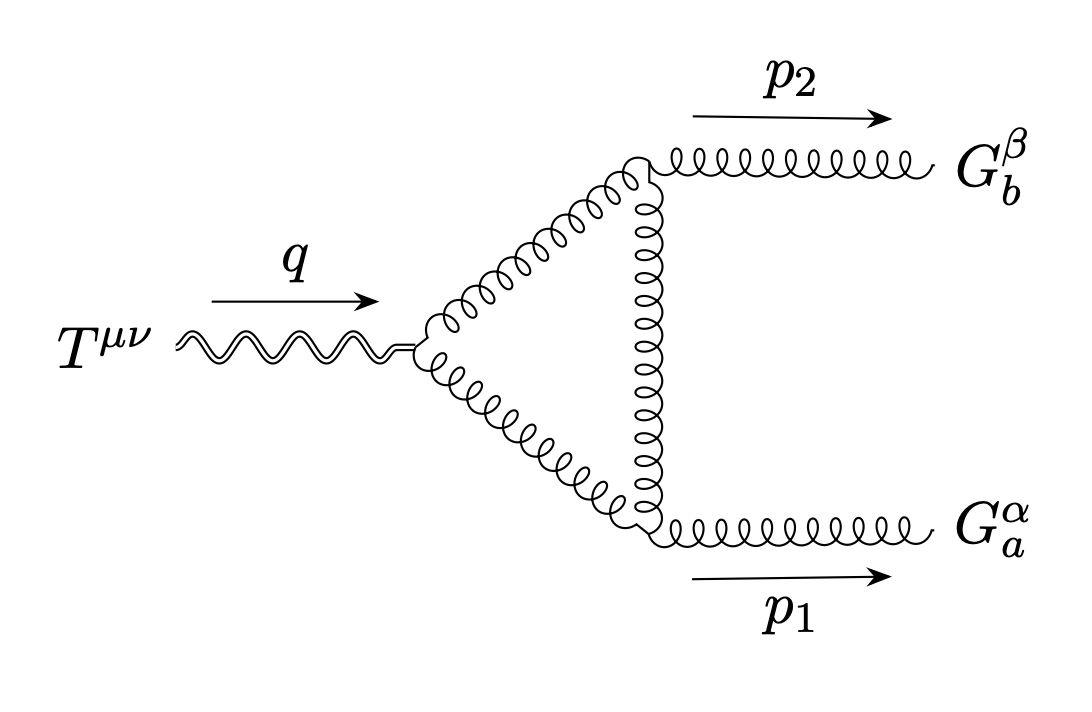}
 \caption{ Typical quark and gluon contributions contributions to the non-Abelian $TJJ$.}
\label{expansion1}
\end{center}
\end{figure}

\section{Tensor decomposition of the $TJJ$ correlator}

The decomposition of the QCD correlator can be presented in a compact form by separating the trace part from the transverse-traceless remainder,
\begin{equation}
\langle T^{\mu\nu}(q)J^{a\alpha}(p_1)J^{b\beta}(p_2)\rangle=
\langle T^{\mu\nu}J^{a\alpha}J^{b\beta}\rangle_{\rm TT}
+
\langle T^{\mu\nu}J^{a\alpha}J^{b\beta}\rangle_{\rm tr},
\label{eq:splitTJJ}
\end{equation}
with $q=p_1+p_2$ being the momentum of the graviton line. Its general expansion 
\begin{equation}
\begin{aligned}
	\Gamma^{\mu\nu\alpha\beta a b}(q,p_1,p_2) & =\langle t^{\mu \nu}(q)  j^{ \alpha a}(p_1) j^{ \beta b}(p_2)\rangle+\langle t^{\mu \nu }(q)j_{loc}^{\alpha  a}(p_1)j^{\beta  b}(p_2)\rangle_g +\langle t^{\mu \nu }(q)j^{\alpha  a}(p_1)j_{loc}^{\beta  b}(p_2)\rangle_g\\
	& \qquad+2 \mathcal{I}^{\mu \nu \rho }(q)\left[\delta_{[\rho}^{\beta} p_{2 \sigma ]}\langle J^{ a\alpha}({p}_1) J^{ b\sigma}(-{p}_1)\rangle+\delta_{[\rho}^{\alpha} p_{1 \sigma]}\langle J^{ b\beta}({p}_2) J^{ a\sigma}(-p_2)\rangle\right] \\
&	+\frac{1}{ 3 \, q^2} \hat\pi^{\mu \nu}(q) \left[\mathcal{A}^{\alpha \beta a b}+\mathcal{B}^{\alpha \beta a b}_g\right]
\end{aligned}
\label{res}
\end{equation}
is organized in terms of transverse traceless contributions such as $\langle t^{\mu \nu}(q)  j^{ \alpha a}(p_1) j^{ \beta b}(p_2)\rangle$; of mixed controbutions such as $\langle t^{\mu \nu }(q)j_{loc}^{\alpha  a}(p_1)j^{\beta  b}(p_2)\rangle_g$ longitudinal in the currents and transverse traceless in the indices of the stress energy tensor; longitudinal ones, proportional to a longitudinal projector $\mathcal{I}^{\mu \nu \rho }$ , and the 
trace anomaly part, proportional to $\mathcal{A}^{\alpha \beta a b}$ given in \eqref{eq:anomAA}. The trace part is also 
accompanied by a term $\mathcal{B}^{\alpha \beta a b}_g$ that vanishes for on-shell gluons. \\
For massive fermions, the trace sector is more involved and takes the schematic form
\begin{equation}
\langle T^{\mu\nu}(q)J^{a\alpha}(p_1)J^{b\beta}(p_2)\rangle_{\rm tr}
=\hat\pi^{\mu\nu}(q)\delta^{ab}\Big[
\Phi_{TJJ}(s,s_1,s_2,m^2)\,u^{\alpha\beta}(p_1,p_2)
+\Phi_2^{\alpha\beta}(s,s_1,s_2,m^2)
\Big],
\label{eq:tracepart}
\end{equation}
where $s=q^2$, $s_1=p_1^2$, $s_2=p_2^2$ and
\begin{equation}
\hat\pi^{\mu\nu}(q)=g^{\mu\nu}q^2 -{q^\mu q^\nu}.
\label{eq:hatpi}
\end{equation}
The first structure in Eq.~\eqref{eq:tracepart} is the one directly associated with the anomaly; the second one is traceful but does not define the conformal-anomaly form factor.

The scalar form factor extracted in \cite{GRSUMRULES-FINALE3MOD} may be written as
\begin{equation}
\Phi_{TJJ}(s,s_1,s_2,m^2)=
\frac{1}{3s}\left[
\frac{g_s^2}{48\pi^2}(11C_A-2n_f)
+\chi_0(s,s_1,s_2,m^2)+\chi_g(s,s_1,s_2)
\right].
\label{eq:PhiTJJ}
\end{equation}
The first term contains the anomaly pole and is directly related to the anomaly. The functions $\chi_0$ and $\chi_g$ collect the massive quark contribution and the purely gluonic correction. Their detailed form is lengthy and can be found in  \cite{Coriano:2024qbr}, but the crucial fact is that they involve scalar two-point and three-point master integrals and therefore generate the branch cuts of the amplitude. Symbolically,
\begin{equation}
\chi_0 = g_s^2 m^4 H_1 C_0(s,s_1,s_2;m^2)
+g_s^2 m^2\sum_{i=2}^{6} H_i\,\mathcal{I}_i(s,s_1,s_2;m^2),
\label{eq:chi0symbolic}
\end{equation}
where the $H_i$ are rational functions of the invariants and the $\mathcal{I}_i$ stand for combinations of $\bar B_0$ and $C_0$ functions, 2-point and 3-point scalar integrals. This representation already makes it evident that, away from the conformal point, i.e. in the presence of massive fermions, the scalar channel is characterized by both pole-like and continuum contributions. A careful analysis uncovers nontrivial patterns of cancellation between poles and cuts in the spectral function of the anomaly form factor, which ultimately give rise to a sum rule for the same quantity. While the derivation is technically involved, the result is exact at one loop. Early indications of analogous sum rules in correlators governed by chiral and conformal anomalies had previously emerged in simpler cases in the past \cite{Horejsi:1997yn}.\\
In the strict massless and on-shell limit, $m\to0$ and $s_1=s_2=0$, Eq.~\eqref{eq:PhiTJJ} simplifies to
\begin{equation}
\Phi_{TJJ}(s,0,0,0)=\frac{g_s^2}{144\pi^2}\frac{11C_A-2n_f}{s},
\label{eq:masslesspole}
\end{equation}
which is the anomaly pole with residue proportional to the QCD beta function. The trace sector then becomes
\begin{equation}
\langle T^{\mu\nu}J^{a\alpha}J^{b\beta}\rangle_{\rm tr}
\xrightarrow[m\to 0]{s_1=s_2=0}
\frac{g_s^2}{144\pi^2}\frac{11C_A-2n_f}{s}
\,\hat\pi^{\mu\nu}(q)\,u^{\alpha\beta}(p_1,p_2)\,\delta^{ab}.
\label{eq:tracepolelimit}
\end{equation}
This is the non-Abelian analogue of the anomaly-pole mechanism familiar from QED \cite{Giannotti:2008cv,Armillis:2009pq}.

\begin{figure}[t]
\begin{center}
\includegraphics[scale=0.6]{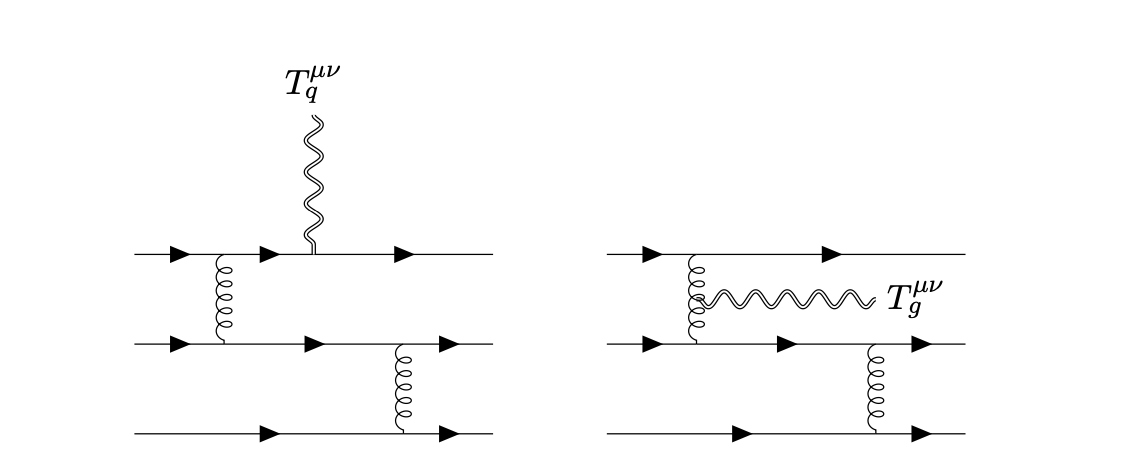}
 \caption{Perturbative expansion of the hard scattering of the GFF in the proton case. The dooubly wiggled line denotes a graviton.}
\label{expansion}
\end{center}
\end{figure}

\section{Spectral density and dispersive sum rule}

The key point of the spectral analysis is that the anomaly form factor admits a dispersion representation in the invariant $s=q^2$,
\begin{equation}
\Phi_{TJJ}(s,s_1,s_2,m^2)=\frac{1}{\pi}\int_{0}^{\infty} \frac{\rho_{TJJ}(s',s_1,s_2,m^2)}{s'-s-i\epsilon}\,ds',
\label{eq:dispPhi}
\end{equation}
with spectral density defined through the discontinuity across the cut,
\begin{equation}
\rho_{TJJ}(s,s_1,s_2,m^2)=\frac{1}{2i}\,\mathrm{Disc}\,\Phi_{TJJ}(s,s_1,s_2,m^2).
\label{eq:rho}
\end{equation}
Because the branch cut starts at the quark threshold $s=4m^2$, the density contains both localized and continuum pieces. Schematically one may write
\begin{align}
\mathrm{Disc}\,\Phi_{TJJ} ={}& c_0(s_1,s_2,m^2)\,\delta(s)
+ c_1(s,s_1,s_2,m^2)\,\theta(s-4m^2) \notag\\
&+ c_2(s_1,s_2,m^2)\,\delta(s-s_1-s_2)
+\cdots ,
\label{eq:discstructure}
\end{align}
where the omitted terms correspond to the cancellation of would-be anomalous thresholds discussed in \cite{GRSUMRULES-FINALE3MOD,Coriano:2024qbr}. The essential result is that the full density satisfies a mass-independent sum rule,
\begin{equation}
\frac{1}{\pi}\int_0^{\infty} ds\, \rho_{TJJ}(s,s_1,s_2,m^2)
=\frac{g_s^2}{144\pi^2}(11C_A-2n_f),
\label{eq:sumrule}
\end{equation}
independently of the mass and of the distribution of spectral strength between pole and continuum.\\
Equation~\eqref{eq:sumrule} is the analogue, for the conformal anomaly, of the sum rules familiar from chiral and chiral-gravitational anomalies \cite{Horejsi:1997yn,Coriano:2025ceu}. It implies that the integrated anomaly strength is distributed over the cut, but it is preserved, even though away from the conformal point the amplitude is no longer exhausted by a simple pole. In particular, the various localized contributions that appear for generic off-shell kinematics, including terms supported at $s=s_1+s_2$, reorganize among themselves in such a way that the final integral remains independent of masses and external virtualities. This cancellation of apparent anomalous-threshold effects is one of the genuinely nontrivial outcomes of the full dispersive analysis and is the reason why the sum rule may be viewed as the cleanest diagnostic of the anomaly channel. As $m\to0$ the continuous part of the density collapses onto the origin and one recovers
\begin{equation}
\rho_{TJJ}(s,s_1,s_2,m^2)\xrightarrow[m\to0]{}
\frac{g_s^2}{144\pi}\,(11C_A-2n_f)\,\delta(s),
\label{eq:deltalimit}
\end{equation}
which reproduces Eq.~\eqref{eq:masslesspole}. In this sense the anomaly reconstructs a dilaton-like interpolating contribution. The ``dilaton'' is not introduced as an elementary degree of freedom but emerges as the massless interpolation selected by the sum rule.
\section{On-shell limit and pole dominance}

The most transparent physical limit is obtained for on-shell gluons, $s_1=s_2=0$. In this case the tensor $u^{\alpha\beta}$ reduces to the familiar gauge-invariant structure for two transverse gauge bosons and the anomaly form factor becomes
\begin{equation}
\Phi_{TJJ}(s,0,0,m^2)=\frac{g_s^2}{144\pi^2}\frac{11C_A-2n_f}{s}
+\frac{g_s^2 n_f m^2}{3s}\,f(s,m^2),
\label{eq:onshellPhi}
\end{equation}
where $f$ is regular at $s=0$. Therefore the massless limit is dominated by the pure pole term. The full trace contribution takes the simple form
\begin{equation}
\Gamma^{\mu\nu\alpha\beta ab}_{\rm tr}(s,0,0)
=\hat\pi^{\mu\nu}(q)u^{\alpha\beta}(p_1,p_2)\delta^{ab}
\left[\frac{g_s^2}{144\pi^2}\frac{11C_A-2n_f}{s}+\mathcal{O}(m^2)\right],
\label{eq:onshelltrace}
\end{equation}
which shows explicitly that the anomaly exchange is nonlocal. The character of the interaction, as shown in 
\cite{GRSUMRULES-FINALE3MOD} is to define a light-cone excitation \cite{GRSUMRULES-FINALE3MOD}.  At the same time, the full $TJJ$ amplitude contains another pole contribution in the traceless sector.
This second structure allows us to identify an additional contribution to the anomaly effective action describing the interaction around the light-cone, of the form
\begin{align}\label{expoleac}
	S_{\text{extra pole}} = \frac{g_s^2}{72 \pi^2} \left(n_f - C_A \right) \int &d^4x \sqrt{-g} \int d^4x' \sqrt{-g'}\, h_{\mu\nu}(x) \Box^{-1}_{x,x'} 
	\nonumber\\&
	\times \left[ 
	3 (\partial^\mu \partial^\nu F^a_{\alpha\beta}) F^{\alpha\beta a}
	+ \frac{1}{4} \left( g^{\mu\nu} \Box - 4 \partial^\mu \partial^\nu \right) F^a_{\alpha\beta} F^{\alpha\beta a}
	\right]_{x'}.
\end{align}
The $n_f$ part was identified in \cite{Giannotti:2008cv} as a possible plasmon mode in QED.
In the non-Abelian case the extra contribution is proportional to $(n_f -C_A)$.  The situation is therefore slightly different from the $AVV$ or $TTJ_5$ cases \cite{Coriano:2023hts,Coriano:2023gxa}, where the residue of the (single) pole present in the anomaly part is sufficient, using the conformal cosntraints, to determine the entire structure of the chiral correlators. 

\section{Relation to anomaly effective actions}

The pole structure discussed above admits a direct interpretation in terms of anomaly-induced effective actions. In a weak-field expansion around flat space, the nonlocal interaction generated by the trace anomaly may be represented schematically as
\begin{equation}
S_{\rm nonlocal}=\frac{\beta(g_s)}{2g_s}\int d^4x\,d^4y\,
R^{(1)}(x)\,\Box^{-1}(x,y)\,[F^a_{\rho\sigma}F^{a\rho\sigma}](y)+\cdots,
\label{eq:nonlocalaction2}
\end{equation}
that can be added to \eqref{expoleac} to define the effective action.
The omitted terms stand for higher-point contributions in the gauge fields and the metric. When Fourier transformed to momentum space, Eq.~\eqref{eq:nonlocalaction2} generates the same scalar $1/q^2$ exchange that appears in Eq.~\eqref{eq:tracepolelimit}. The effective action is therefore not merely a formal rewriting of the perturbative result; it captures the same long-range kinematics singled out by the sum rule in the conformal limit.\\
It is important, however, not to over-interpret this action. The object interpolated by the anomaly pole is not a standard elementary scalar. The effective nonlocal vertex is defined by a constrained sector of the three-point function and is tied to the sum rule obeyed by the spectral density. In this respect it is more accurate to view the anomaly pole as an emergent collective interpolation, analogous in spirit to the effective pole descriptions that appear in other light-cone-dominated regimes.\\
This observation also clarifies the distinction between the present $TJJ$ analysis and the study of higher stress energy tensor correlators such as $TTT$. The latter probe the full gravitational sector of the anomaly and involve a substantially richer set of tensor structures \cite{Coriano:2017mux,Coriano:2018bsy}. By contrast, the $TJJ$ correlator isolates the coupling of one stress energy tensor to two gauge currents and is the natural object for the partonic discussion of GFFs, being the anomaly dependent on the gauge fields. The two problems are certainly related, but they are conceptually distinct. The first is mainly concerned with the gauge structure, while the second involves the full nonlinear realization of the conformal anomaly in a gravitational framework, with the inclusion of additional Weyl-invariant terms such as $C^2$, the square of the Weyl tensor, and $E$, the Euler--Poincar\'e density, or Gauss--Bonnet term, absent in the GFF case.
\section{Physical interpretation and relation to hadronic observables}

The partonic analysis summarized above does not yet provide the hadronic GFFs directly. Rather, it identifies which short-distance structures survive in the light-cone limit and therefore may feed the hadronic form factors once convoluted with the relevant nonperturbative matrix elements. In this sense the anomaly analysis is not separate from GFF phenomenology. It is a microscopic characterization of the kernel that appears in the partonic description of stress-tensor insertions.\\
At the same time, one should be precise about the physical meaning of GFFs. They do not measure dynamical gravity. They measure matrix elements of the QCD stress energy tensor, with the metric treated as an external source. To probe genuine gravitational dynamics one would have to move to higher stress-tensor correlators such as $TTT$, which, as just mentioned, project onto other sectors of the conformal anomaly and lie beyond the immediate scope of the hadronic problem \cite{Coriano:2017mux,Coriano:2018bsy}. The present discussion is therefore best viewed as a bridge between two problems: the extraction of hadronic structure from exclusive QCD amplitudes and the field-theoretic interpretation of anomaly-induced exchanges in perturbation theory. In this sense the $TJJ$ correlator occupies a privileged intermediate position. It is simpler than the full purely gravitational sector, but already rich enough to display the emergence of a scalar anomaly channel, its dispersive sum rule and its interplay with the ordinary traceless tensor exchange relevant for partonic scattering.\\
The main conceptual lesson is that an anomaly pole should not be interpreted naively as an asymptotic scalar particle. Its physical role is that of an interpolating contribution whose existence is protected by the sum rule and whose residue is determined by the anomaly. Away from the conformal point the pole shares its spectral weight with the continuum; in the conformal light-cone limit the continuum collapses and the pole fully saturates the sum rule. This is the sense in which one may speak of a dilaton-like exchange in perturbative QCD.

\section{Why the sum rule matters for GFFs}

From the viewpoint of exclusive QCD processes, the sum rule is important because it singles out a universal part of the hard kernel entering stress-tensor insertions. To see the point, one may write the partonic contribution to a generic GFF amplitude in the symbolic form
\begin{equation}
\mathcal{M}_{\rm GFF}\sim \int [dx_i] [dy_i]\, \Psi^{\ast}_{H'}(y_i,\mu)
\,\mathcal{K}_{TJJ}(x_i,y_i,Q^2,t;\mu)
\,\Psi_H(x_i,\mu),
\label{eq:hardkernelGFF}
\end{equation}
where $\Psi_H$ denotes a hadronic light-cone wave function and $\mathcal{K}_{TJJ}$ the hard kernel generated by the insertion of the energy-momentum tensor in the partonic subprocess. The anomaly analysis implies that the scalar part of $\mathcal{K}_{TJJ}$ contains a universal light-cone component controlled by Eqs.~\eqref{eq:PhiTJJ} and \eqref{eq:sumrule}. This does not mean that hadronic GFFs are themselves fixed by the anomaly, but it does imply that one identifiable sector of the hard amplitude has a protected dispersive normalization.

A related point concerns the asymptotic large-$Q^2$ behavior. Since the anomaly pole is associated with the scalar exchange $1/q^2$, its effect survives in the same kinematic region in which collinear factorization is applicable. Consequently, in a perturbative treatment of GFFs at large momentum transfer, the anomaly-mediated contribution is not power suppressed relative to the natural graviton insertion. What distinguishes it is not its scaling but its tensorial projection and its protected spectral normalization. This is the precise sense in which the anomaly sector is dynamically relevant for the light-cone description of gravitational form factors.

Finally, the sum rule gives a principled criterion for separating the anomaly contribution from ordinary explicit scale breaking. Terms proportional to the quark mass do contribute to the trace, but they do not possess a protected mass-independent integral analogous to Eq.~\eqref{eq:sumrule}. The anomaly form factor is therefore the unique scalar structure whose integrated spectral weight is rigid under deformations away from the conformal point. This rigidity is the reason the corresponding exchange can be tracked continuously from generic off-shell kinematics to the conformal on-shell pole.

\section{Conclusions}

The analysis summarized here combines three themes that are usually discussed separately: the extraction of gravitational form factors from exclusive QCD processes, the momentum-space conformal analysis of stress-tensor correlators, and the dispersive interpretation of anomaly-induced interactions. Their interplay leads to a coherent framework in which the partonic light-cone structure of hadronic amplitudes is connected to the trace anomaly of QCD through the $TJJ$ correlator.\\
The first main ingredient is the standard factorization program for DVCS and related processes, through which moments of generalized parton distributions provide access to the proton gravitational form factors and to Ji's sum rule \cite{Coriano:2024qbr}. The second is the embedding of QCD in curved spacetime, which allows one to define and analyze the relevant stress-tensor correlators in a systematic way. The third is the identification of a distinguished scalar sector in the tensor decomposition of the $TJJ$ three-point function, together with the demonstration that its spectral density obeys a mass-independent sum rule \cite{Coriano:2024qbr,GRSUMRULES-FINALE3MOD}. Sum rules of these types, in the chiral case, could also clarify some of the issues debated on the chiral anomaly and the gluon polarization 
\cite{Castelli:2024eza,Tarasov:2021yll,Bhattacharya:2023wvy}.\\
This sum rule suggests the presence of an effective dilaton-like interpolating channel, reconstructed from the continuum and collapsing to a pole at the conformal point. The terminology should be understood in an interpolating sense: one is not introducing a fundamental scalar into the microscopic Lagrangian, but identifying a protected scalar exchange selected by the anomaly and reconstructed dispersively from the full correlator. In the light-cone limit, the anomaly contribution and an additional traceless term (the plasmon mode) both play a role, motivating an effective-action description of the short-distance interaction. \\
A natural next step is the extension of this framework from the partonic to the hadronic level in a fully systematic way. Such a development would sharpen the connection with phenomenology at facilities such as the Electron-Ion Collider and would clarify how the perturbative anomaly dynamics is encoded in measurable hadronic observables. More broadly, the same tools may prove useful in other contexts in which conformal anomalies, light-cone dominance and effective scalar exchanges play a central role.

\acknowledgments
This work is partially supported by INFN, inziativa specifica {\em QG-sky}, by the the grant PRIN 2022BP52A MUR "The Holographic Universe for all Lambdas" Lecce-Naples, and by the European Union, Next Generation EU, PNRR project "National Centre for HPC, Big Data and Quantum Computing", project code CN00000013. 

\end{document}